\documentclass[twocolumn,english,superscriptaddress,floatfix]{revtex4}
\usepackage[T1]{fontenc}
\usepackage[latin9]{inputenc}
\usepackage{color}
\usepackage{babel}
\usepackage{amsmath}
\usepackage{amssymb}
\usepackage{graphicx}
\usepackage{esint}
\usepackage[unicode=true,
 bookmarks=false,
 breaklinks=true,pdfborder={0 0 1},colorlinks=true]
 {hyperref}
\hypersetup{
 pdfcreator={},pdfproducer={LaTeX with hyperref},linkcolor=blue,anchorcolor=blue,citecolor=blue,filecolor=red,menucolor=red,pagecolor=red,urlcolor=blue,pdfstartview=FitV,pdfhighlight=/I,pdfpagelayout=OneColumn,hypertexnames=true}
\usepackage{breakurl}

\makeatletter
\@ifundefined{textcolor}{}
{%
 \definecolor{BLACK}{gray}{0}
 \definecolor{WHITE}{gray}{1}
 \definecolor{RED}{rgb}{1,0,0}
 \definecolor{GREEN}{rgb}{0,1,0}
 \definecolor{BLUE}{rgb}{0,0,1}
 \definecolor{CYAN}{cmyk}{1,0,0,0}
 \definecolor{MAGENTA}{cmyk}{0,1,0,0}
 \definecolor{YELLOW}{cmyk}{0,0,1,0}
}

\usepackage{babel}
\usepackage{babel}

\newcommand{\be}{\begin{equation}}
\newcommand{\ee}{\end{equation}}
\newcommand{\bea}{\begin{eqnarray}}
\newcommand{\eea}{\end{eqnarray}}
\newcommand{\bse}{\begin{subequations}}
\newcommand{\ese}{\end{subequations}}

\definecolor{d_red}{cmyk}{0.00, 0.81, 1.00, 0.27}
\definecolor{d_orange}{cmyk}{0.00, 0.33, 1.00, 0.00}
\definecolor{d_blue}{cmyk}{0.78, 0.47, 0.00, 0.20}
\definecolor{d_lgreen}{cmyk}{0.07, 0.00, 0.79, 0.29}
\definecolor{d_green}{cmyk}{0.66, 0.00, 0.71, 0.56}
\definecolor{d_blue}{cmyk}{0.78, 0.47, 0.00, 0.20}
\definecolor{d_dblue}{cmyk}{0.91, 0.79, 0.00, 0.22}
\definecolor{d_pink}{cmyk}{0.0, 0.79, 0.37, 0.29}
\definecolor{d_purple}{cmyk}{0.16, 0.54, 0.00, 0.70}
\definecolor{d_paleblue}{cmyk}{0.669, 0.338, 0.00, 0.373}
\definecolor{d_dpaleblue}{cmyk}{0.441, 0.290, 0.00, 0.580}
\definecolor{d_brown}{cmyk}{0.0, 0.490, 0.930, 0.350}
\definecolor{d_turquoise}{cmyk}{0.630, 0.04, 0.0, 0.440}
\definecolor{KIT-green}{RGB}{0, 150,130}
\definecolor{KIT-blue}{RGB}{70,100,170}

\def\bmx{\begin{pmatrix}}
\def\emx{\end{pmatrix}}

\usepackage[figure,table]{hypcap}
\usepackage{dsfont}

\makeatother

\begin{document}

\title{Vestigial nematic order and superconductivity in the doped topological
insulator Cu$_{x}$Bi$_{2}$Se$_{3}$ }

\author{Matthias Hecker }

\affiliation{Institut für Theorie der Kondensierten Materie, Karlsruher Institut
für Technologie, 76131 Karlsruhe, Germany}

\affiliation{Institut für Festkörperphysik, Karlsruher Institut für Technologie,
76344 Karlsruhe, Germany}

\author{Jörg Schmalian}

\affiliation{Institut für Theorie der Kondensierten Materie, Karlsruher Institut
für Technologie, 76131 Karlsruhe, Germany}

\affiliation{Institut für Festkörperphysik, Karlsruher Institut für Technologie,
76344 Karlsruhe, Germany}

\date{\today }
\begin{abstract}
If the topological insulator Bi$_{2}$Se$_{3}$ is doped with electrons,
superconductivity with $T_{{\rm c}}\approx3-4\:{\rm K}$ emerges for
a low density of carriers ($n\approx10^{20}{\rm cm}^{-3}$) and with
a small ratio of the superconducting coherence length and Fermi wave
length: $\xi/\lambda_{F}\approx2\cdots4$. These values make fluctuations
of the superconducting order parameter increasingly important, to
the extend that the $T_{c}$-value is surprisingly large. Strong spin-orbit
interaction led to the proposal of an odd-parity pairing state. This
begs the question of the nature of the transition in an unconventional
superconductor with strong pairing fluctuations. We show that for
a multi-component order parameter, these fluctuations give rise to
a nematic phase at $T_{{\rm nem}}>T_{c}$. Below $T_{c}$ several
experiments demonstrated a rotational symmetry breaking where the
Cooper pair wave function is locked to the lattice. Our theory shows
that this rotational symmetry breaking, as vestige of the superconducting
state, already occurs above $T_{c}$. The nematic phase is characterized
by vanishing off-diagonal long range order, yet with anisotropic superconducting
fluctuations. It can be identified through direction-dependent para-conductivity,
lattice softening, and an enhanced Raman response in the $E_{g}$
symmetry channel. In addition, nematic order partially avoids the
usual fluctuation suppression of $T_{c}$. 
\end{abstract}
\maketitle
The electron doped topological insulator Bi$_{2}$Se$_{3}$ has been
reported to exhibit a low carrier density ($n\approx10^{20}{\rm cm}^{-3}$),
together with a small ratio $\xi/\lambda_{F}\approx2\cdots4$ \cite{Hor2010,Kriener2011,Kriener2011-1}.
Recently, NMR Knight-shift measurements \cite{Matano2016} and measurements
of the angular-dependent specific heat in the magnetic field \cite{Yonezawa2017}
revealed spontaneous symmetry breaking of the superconducting state
in addition to the global $U\left(1\right)$-symmetry. The threefold
symmetry of the underlying lattice is broken. Similar \emph{nematic}
superconductivity was observed in Sr$_{x}$Bi$_{2}$Se$_{3}$ \cite{Pan2016,Du2017}
and in the closely related Nb-doped Bi$_{2}$Se$_{3}$ \cite{Asaba2017,Shen2017}.
Early on, Fu and Berg made the proposal that Cu$_{x}$Bi$_{2}$Se$_{3}$
may have an odd-parity two-component superconducting order parameter
\cite{Fu2010}. Rotational symmetry breaking below $T_{c}$ is then
a possible consequence of this pairing state which is, by symmetry,
in the $E_{u}$ representation of the point group $D_{3d}$, see Ref.
\cite{Fu2010,Fu2014,Venderbos2016,Venderbos2017}. This is the representation
that transforms like the in-plane coordinates $\mathbf{x}=\left(x,y\right)$.

\begin{figure}
\includegraphics[scale=0.4]{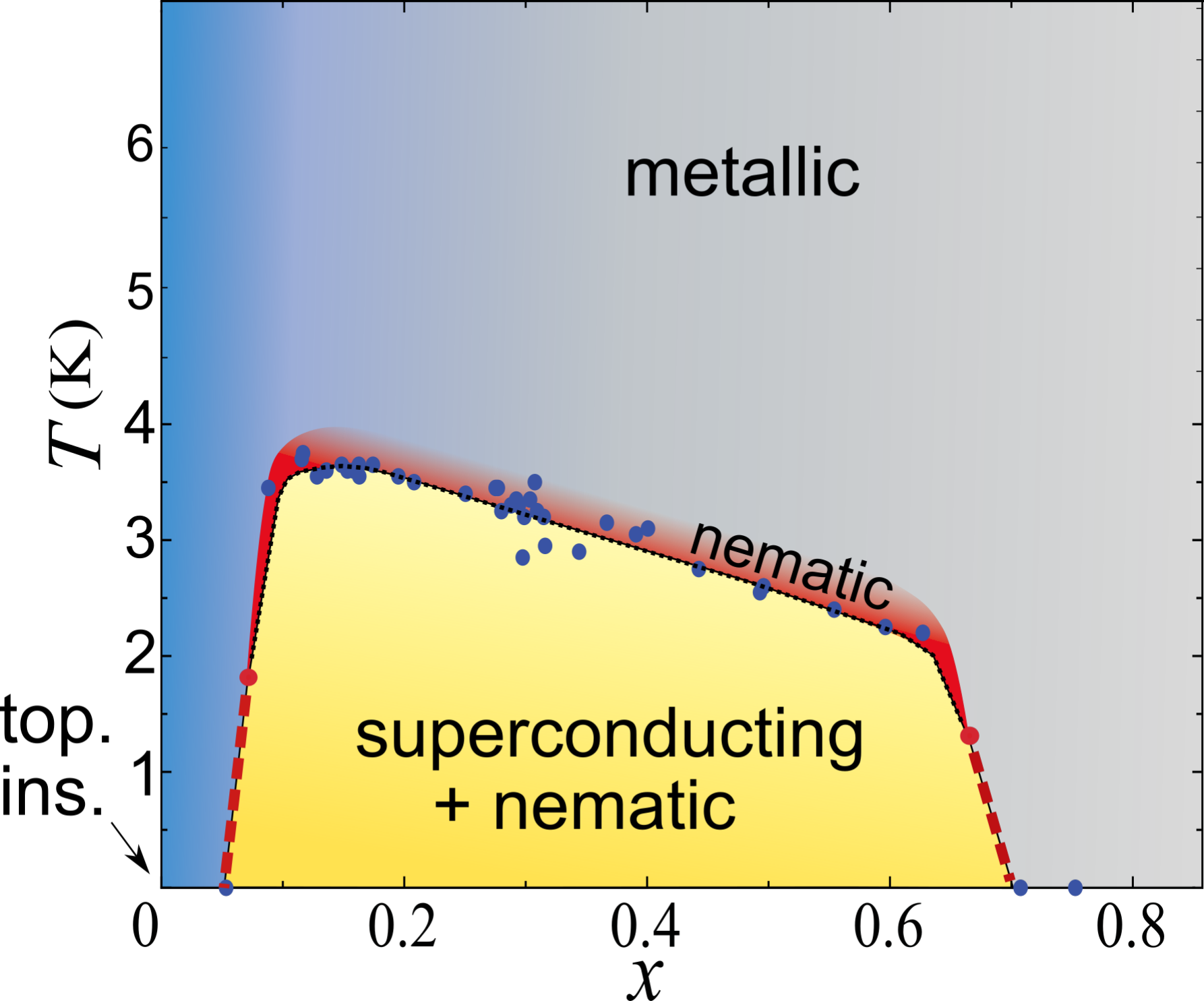}

\protect\protect\protect\caption{Schematic phase diagram of Cu$_{x}$Bi$_{2}$Se$_{3}$ with data points
(blue bullets) taken from \cite{Kriener2011-1}. We predict a purely
nematic phase above the superconducting phase (indicated in red),
where superconducting fluctuations create an ordered state that breaks
the threefold rotational symmetry. Following Ref. \cite{Fernandes2012},
we expect for low temperatures tricritical points (red bullets) below
which the transitions should be joint first order. \label{fig:phase_diag}}
\end{figure}

In this paper we show that superconducting fluctuations induce a phase
transition to a nematic state. We find that these fluctuations either
give rise to a nematic phase transition at $T_{{\rm nem}}>T_{{\rm c}}$
or drive the superconducting transition weakly first order. Our quantitative
analysis prefers the former scenario, where nematicity is a vestigial
precursor phase of superconductivity. This is due to the pronounced
two-dimensional electronic structure seen in ARPES measurements \cite{Lahoud2012}
that is induced by Cu-intercalation. In distinction to the usual expectation
where fluctuations suppress $T_{c}$, we find that nematic order largely
off-sets this suppression, i.e. strengthens pairing compared to the
case without nematic order. In the nematic state, the overall superconducting
phase averages out to zero, yet the relative orientation of the two
components of the Cooper pair field condenses in a long ranged ordered
state with broken $Z_{3}$ or three-states Potts model symmetry at
$T_{{\rm nem}}$. Superconductivity sets in at a temperature slightly
below $T_{{\rm nem}}$. The resulting phase diagram for doped Bi$_{2}$Se$_{3}$
is sketched in Fig.\ref{fig:phase_diag}. Because of the locking of
the Cooper pair wave function to the lattice, the\textcolor{red}{{}
}elastic constant $c_{E_{g}}$ together with the sound velocity along
certain high-symmetry directions are reduced at the upper temperature
$T_{{\rm nem}}$. As the nematic transition of a clean system turns
out to be weakly first order, the elastic constant will however not
completely vanish. Weak disorder changes the transition to become
second order giving rise to a vanishing elastic constant. The nematic
state above $T_{c}$ can also be identified through anisotropic paraconductvity
$T_{{\rm c}}<T<T_{{\rm nem}}$. We determine this anisotropy from
the fluctuation spectrum of the Cooper pair field. The nematic order
discussed here has several parallels to spin-induced Ising nematic
order above a striped magnetic state of the iron-based superconductors
\cite{Fang2008,Xu2008,Fernandes2010,Fernandes2012} or to time-reversal
symmetry breaking proposed for chiral superconductors in the context
of SrRuO$_{4}$ \cite{Fischer2016}, revealing the universality of
the underlying principle of composite or intertwined order \cite{Fradkin2015}.

Before we discuss the details of our analysis we summarize the key
idea of this paper. The low energy Hamiltonian that describes the
superconducting state of doped Bi$_{2}$Se$_{3}$ in the band basis
is of the form 

\begin{align}
H & \approx\sum_{\mathbf{k}s}\varepsilon_{\mathbf{k}}\psi_{\mathbf{k},s}^{\dagger}\psi_{\mathbf{k},s}+\sum_{\mathbf{k}}\left[\psi_{\mathbf{k}}^{\dagger}\left(\mathbf{d}_{\mathbf{k}}\cdot\boldsymbol{\tilde{\sigma}}\right)i\tilde{\sigma}^{y}\psi_{-\mathbf{k}}^{\dagger}+h.c.\right],\label{eq:Ham_band}
\end{align}
with the pseudo spin index $s=\left\{ 1,2\right\} $ and $\psi_{\mathbf{k}}^{\dagger}=\left(\psi_{\mathbf{k},1}^{\dagger},\psi_{\mathbf{k},2}^{\dagger}\right)$
denoting fermionic creation operators in the conduction band (cf.
Methods section for details). The pairing is given by the $\mathbf{d}_{\mathbf{k}}$-vector
as

\begin{align}
\mathbf{d}_{\mathbf{k}} & =\Delta^{\left(x\right)}\left(k_{z}\mathbf{e}_{x}-\frac{v_{0}}{v_{z}}k_{x}\mathbf{e}_{z}\right)+\Delta^{\left(y\right)}\left(k_{z}\mathbf{e}_{y}-\frac{v_{0}}{v_{z}}k_{y}\mathbf{e}_{z}\right),\label{eq:d_vec}
\end{align}
where $\mathbf{\Delta}=\left(\Delta^{\left(x\right)},\Delta^{\left(y\right)}\right)^{T}$
form a two-component order parameter in the $E_{u}$ representation.
The broken rotation symmetry below $T_{c}$ naturally implies that
this is the appropriate pairing state. Within the $D_{3d}$ point
group,\textcolor{red}{{} }the only alternative would be pairing in the
even-parity state $E_{g}$. This corresponds to $\left(d_{x^{2}-y^{2}}+d_{yz},\, d_{xy}+d_{xz}\right)$
superconductivity. Most of our analysis would proceed without changes
if this were the case. Nematic order can generally be characterized
in terms of a symmetric trace-less second rank tensor. The nematic
tensor in our problem is

\begin{equation}
\hat{q}_{\alpha\beta}=-\sum_{\mu,\nu}\hat{\boldsymbol{\tau}}_{\mu\nu}\cdot\hat{\boldsymbol{\tau}}_{\alpha\beta}\Delta^{\left(\mu\right)*}\Delta^{\left(\nu\right)},\label{eq:bilin}
\end{equation}
where we use the Pauli matrices $\hat{\boldsymbol{\tau}}=\left(\hat{\tau}^{z},\hat{\tau}^{x}\right)$.
Throughout this paper, we use hat symbols for $2\times2$ matrices
and bold symbols for vectors. The expectation value $\left\langle \Delta^{\left(\mu\right)*}\Delta^{\left(\nu\right)}\right\rangle $
measures Cooper pair correlations. While such an expectation value
does not break an additional symmetry in the case of a single-component
pairing state, we show below that $\left\langle \hat{q}_{\alpha\beta}\right\rangle \neq0$
breaks another symmetry. For example, $\left\langle \hat{q}_{xx}\right\rangle \neq0$
implies $\left\langle \Delta^{\left(x\right)*}\Delta^{\left(x\right)}\right\rangle \neq\left\langle \Delta^{\left(y\right)*}\Delta^{\left(y\right)}\right\rangle $,
while in the high temperature phase both expectation values are equal.
We show that the additional symmetry is separately broken at a distinct
temperature. The emerging nematic phase is not superconducting but
induced by superconducting fluctuations. It is a vestige of the superconducting
phase. This is only possible because $\mathbf{\boldsymbol{\Delta}}$
is a two-component order parameter, i.e. the irreducible representation
$E_{u}$ has dimensionality two. To be precise, $\left\langle \hat{q}_{\alpha\beta}\right\rangle $
transforms according to the representation $E_{g}$ of the point group.
Since $E_{u}\otimes E_{u}=A_{1g}\oplus A_{2g}\oplus E_{g}$ such a
composite order, made up of a bilinear combination, is indeed allowed.
Since $E_{u}\otimes E_{u}=E_{g}\otimes E_{g}$ the analysis of nematic
order does not change for even-parity multi-component superconductivity.
The other non-trivial bilinear form is $q^{y}\equiv\sum_{\mu\nu}\Delta^{\left(\mu\right)*}\hat{\tau}_{\mu\nu}^{y}\Delta^{\left(\nu\right)}$,
which transforms under $A_{2g}$ and breaks time reversal symmetry.
In what follows we focus on the nematic order parameter $\left\langle \hat{q}_{\alpha\beta}\right\rangle $.
It takes the general form

\begin{eqnarray}
\left\langle \hat{q}_{\alpha\beta}\right\rangle  & = & -2q_{0}\left(n_{\alpha}n_{\beta}-\frac{1}{2}\delta_{\alpha\beta}\right).\label{eq:nem_op}
\end{eqnarray}
The amplitude $q_{0}$ sets in at the nematic transition temperature
$T_{{\rm nem}}$. The unit vector $\mathbf{n}=\left(\cos\theta,\sin\theta\right)^{T}$
is the director of the nematic state that determines the eventual
orientation of the superconducting order parameter 
\begin{equation}
\mathbf{\Delta}=\Delta_{0}\left(\cos\theta,\sin\theta\right)^{T}.\label{eq:rot solution-1}
\end{equation}
Thus, the superconducting Cooper pair field acts as nematogen that
enables a rotational symmetry breaking, even without superconducting
long-range order. Finally, the lattice symmetry of Cu$_{x}$Bi$_{2}$Se$_{3}$
allows for three distinct values of the angle $\theta=\left\{ 0,\frac{\pi}{3},\frac{2\pi}{3}\right\} $
(cf. Fig.\ref{fig:values_OPs}). The statistical mechanics of the
nematic state then corresponds to a three state Potts model.

\section*{Results}

\paragraph{Collective nematic fluctuations.}

Our starting point is the well established microscopic Hamiltonian
(\ref{eq:h_k-1}) for Bi$_{2}$Se$_{3}$, yet, for the sake of clarity
we refer to the Methods section for a discussion. It leads to the
Ginzburg-Landau expansion valid in the vicinity of the superconducting
phase transition. In terms of the two-component order parameter $\mathbf{\Delta}=\left(\Delta^{\left(x\right)},\Delta^{\left(y\right)}\right)^{T}$
the action reads:

\begin{align}
\mathcal{S} & =r_{0}\int_{x}\mathbf{\Delta}^{\dagger}\mathbf{\Delta}\;\;+\;\;\mathcal{S}^{{\rm grad}}\nonumber \\
 & \quad+\int_{x}\left(u\,\left(\mathbf{\Delta}^{\dagger}\mathbf{\Delta}\right)^{2}+v\,\left(\mathbf{\Delta}^{\dagger}\hat{\tau}_{y}\mathbf{\Delta}\right)^{2}\right)\;,\label{eq:eq:action_delta}
\end{align}
with $r_{0}=\frac{1}{g}-\rho_{F}\log\frac{\omega_{0}}{T}$, where
$g$ and $\omega_{0}$ are the strength and characteristic energy
of the pairing interaction, and $\rho_{F}$ the density of states
at the Fermi level, respectively. The gradient term is in momentum
space given as 
\begin{equation}
\mathcal{S}^{{\rm grad}}=\int_{p}\mathbf{\Delta}_{\mathbf{p}}^{\dagger}\left(m_{0}\left(\mathbf{p}\right)\hat{\tau}_{0}+\mathbf{m}\left(\mathbf{p}\right)\cdot\hat{\mathbf{\boldsymbol{\tau}}}\right)\mathbf{\Delta}_{\mathbf{p}}\label{eq:gradient_action}
\end{equation}
with $m_{0}\left(\mathbf{p}\right)=d_{\parallel}\left(p_{x}^{2}+p_{y}^{2}\right)+d_{z}\, p_{z}^{2}$,
$m_{1}\left(\mathbf{p}\right)=d'\left(p_{x}^{2}-p_{y}^{2}\right)+\bar{d}\, p_{y}p_{z}$,
and $m_{2}\left(\mathbf{p}\right)=2d'p_{x}p_{y}+\bar{d}\, p_{x}p_{z}$,
and characterized by four parameters \cite{Venderbos2017}. We use
the shorthand notation $\int_{p}\cdots=T\int\frac{d^{3}\mathbf{p}}{\left(2\pi\right)^{3}}\cdots$.
From the microscopic Hamiltonian one can determine the coefficients
of the Ginzburg-Landau expansion at weak coupling $\lambda=g\rho_{F}\ll1$.
This analysis yields $u>0$ and $v>0$, in full agreement with earlier
calculations \cite{Venderbos2016}. Crucial for our analysis is however
not the applicability of this expansion, but only the signs of $u$
and $v$. The key implication of the positive sign of $v$ is that
the superconducting order parameter is time reversal symmetric and
can be written in the form of Eqn. (\ref{eq:rot solution-1}). If
$v<0$ we would find time-reversal symmetry breaking and an analyis
analogous to ours leads to a vestigial order parameter $\left\langle q^{y}\right\rangle $.
The regime $v<0$ was predicted in Ref.\cite{Chirolli2018} for thin
layers of doped Bi$_{2}$Se$_{3}$. Thus, no matter what the sign
of $v$, one always has an accompanying symmetry breaking. This is
true for all crystalline symmetries that allow for multi-component
Cooper pair fields. For a superconducting order parameter, transforming
according to a higher dimensional irreducible representation $\Gamma$,
the product representation $\Gamma^{*}\otimes\Gamma$ contains non-trivial
irreducible represenations. For all point groups relevant to periodic
systems these give rise to vestigial order parameters. Thus, there
is no direct second-order superconducting phase transition for a multi-component
Cooper pair field. Either there is a vestigial phase above $T_{c}$,
or the transition is of first order due to the coupling to the vestigial
order parameter.

\begin{figure}
\includegraphics[scale=0.4]{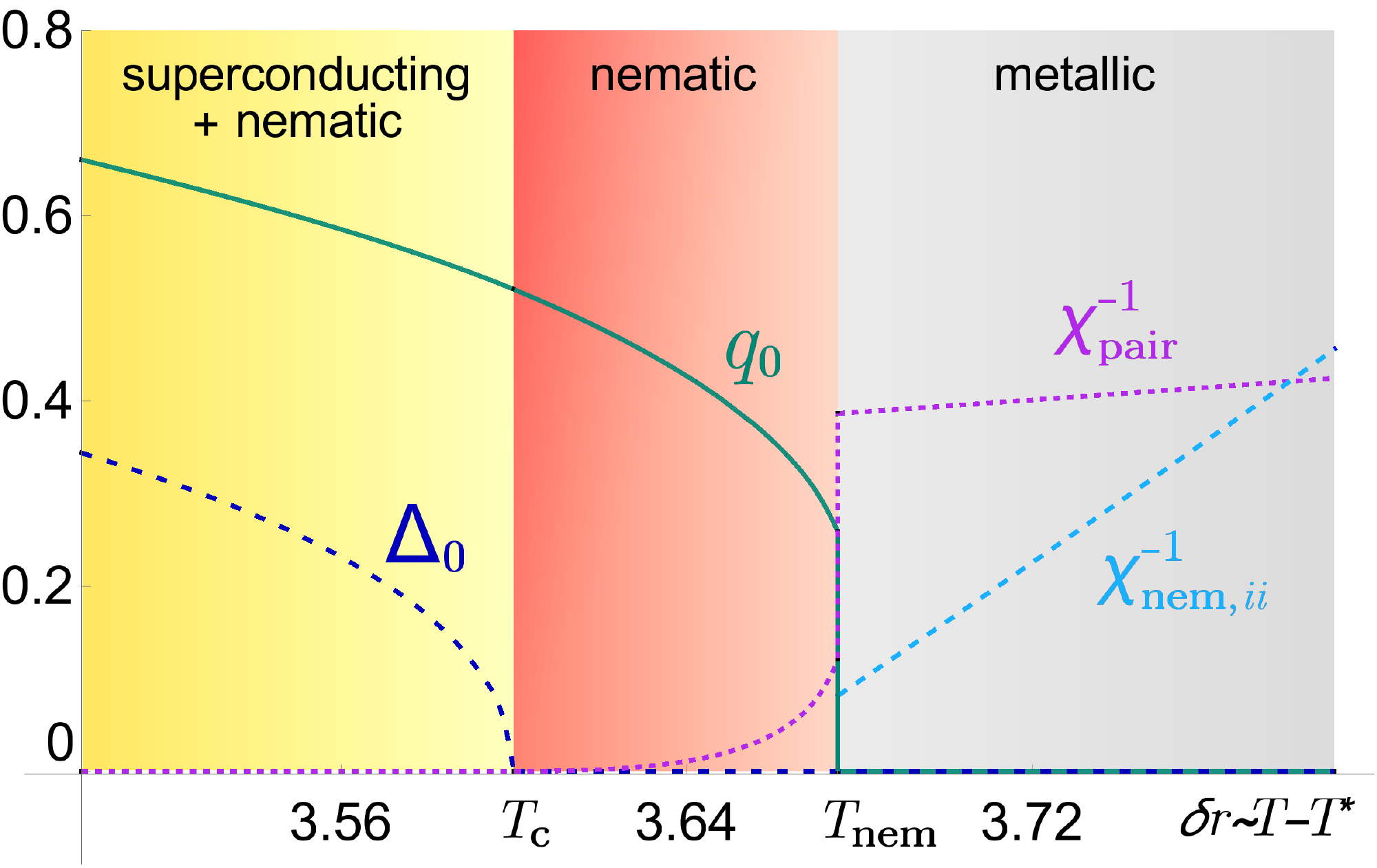}

\protect\protect\protect\caption{The calculated temperature dependence of the nematic ($q_{0}$) and
the superconducting ($\Delta_{0}$) order parameters. The temperature
axis is given by $\delta r=r_{0}-r_{0}^{*}\propto T-T^{*}$, where
$T^{*}$ is the transition temperature without nematic order. The
positive value of $T_{c}$ demonstrates its enhancement due to nematicity
as compared to the case with fluctuations but without nematic order.
We find a first order nematic phase transition at $T=T_{{\rm nem}}$,
followed by a second order superconducting phase transition at $T=T_{c}$.
We also depict the inverse uniform pairing susceptibility $\chi_{{\rm pair}}^{-1}$,
which experiences a sudden drop at $T_{{\rm nem}}$ before it vanishes
at $T_{c}$. Finally, we show the inverse nematic susceptibility $\chi_{{\rm nem}}^{-1}$
that reflects nematic fluctuations above $T_{{\rm nem}}$. Plotted
are the dimensionless quantities $\Delta_{0}\sqrt{d_{\parallel}V_{0}^{2}/(T_{c}V\tilde{v})}$,
$q_{0}/(2d_{\parallel}\tilde{v})$, $\chi_{{\rm pair}}^{-1}/(d_{\parallel}\tilde{v})$
and $\delta r/(d_{\parallel}\tilde{v})$ with definitions and parameter
values given in the Methods section.\label{fig:OP_fig} }
\end{figure}

For an analysis of fluctuation effects and the description of the
nematic ordering, it is efficient to express the interaction in terms
of the quadrupolar tensor $\hat{q}_{\alpha\beta}$ of Eqn. (\ref{eq:bilin}):
\begin{equation}
\mathcal{S}^{\left(4\right)}=\frac{u'}{2}\int_{x}{\rm tr}\left(\hat{r}\hat{r}\right)-\frac{v}{2}\int_{x}{\rm tr}\left(\hat{q}\hat{q}\right)\;,
\end{equation}
where we introduced $u'=u+v$ and $\hat{r}=\left(\mathbf{\Delta}^{\dagger}\mathbf{\Delta}\right)\hat{\tau}_{0}$.
We decouple the two terms in $\mathcal{S}^{\left(4\right)}$ via Hubbard-Stratonovich
transformations, e.g. $\int D\hat{Q}\, e^{-\frac{1}{8v}{\rm tr}\left(\hat{Q}\hat{Q}\right)-\frac{1}{2}{\rm {\rm tr}\left(\hat{Q}\hat{q}\right)}}\propto e^{\frac{v}{2}{\rm tr}\left(\hat{q}\hat{q}\right)}$,
and obtain

\begin{align}
\mathcal{S} & =\frac{1}{4}\int_{x}\left(\frac{1}{v}\,\mathbf{Q}\cdot\mathbf{Q}-\frac{1}{u'}\, R^{2}\!\right)+\int_{p}\mathbf{\Delta}_{\mathbf{p}}^{\dagger}\,\hat{{\cal \chi}}_{\mathbf{p}}^{-1}\mathbf{\Delta}_{\mathbf{p}}\;,\label{eq:action3}
\end{align}
with the pairing susceptibility 
\begin{equation}
\hat{\chi}_{\mathbf{p}}^{-1}=\left(r_{0}+R+m_{0}\left(\mathbf{p}\right)\right)\hat{\tau}_{0}+\left(\mathbf{Q}+\mathbf{m}\left(\mathbf{p}\right)\right)\cdot\hat{\mathbf{\boldsymbol{\tau}}}\,.\label{eq:chip}
\end{equation}
Here, we have expanded the matrices $\hat{R}=R\,\hat{\tau}_{0}$ and
$\hat{Q}=\mathbf{Q}\cdot\hat{\mathbf{\boldsymbol{\tau}}}$ in terms
of the Pauli matrices with $\mathbf{Q}=\left(Q_{1},Q_{2}\right)^{T}$.
Next, the superconducting order parameter fluctuations are integrated
out in both regimes, $T<T_{c}$ and $T>T_{c}$. To this end, we include
Gaussian fluctuations of the pairing field, treated formally within
a large-$N$ expansion of the vector field $\boldsymbol{\Delta}$.
We also allow for superconducting symmetry breaking with the condensed
pairing field $\mathbf{\Delta}_{0}=\left(\Delta_{0}^{\left(x\right)},\Delta_{0}^{\left(y\right)}\right)$
where $\Delta_{0}^{\left(x,y\right)}\in\mathbb{R}$:

\begin{align}
\mathcal{S} & =\frac{1}{4}\int_{x}\left(\frac{1}{v}\,\mathbf{Q}\cdot\mathbf{Q}-\frac{1}{u'}\, R^{2}\!\right)+\mathbf{\Delta}_{0}^{T}\,\hat{{\cal \chi}}_{0}^{-1}\mathbf{\Delta}_{0}\nonumber \\
 & \quad+\int_{p}{\rm tr}\log\hat{{\cal \chi}}_{\mathbf{p}}^{-1}\;.\label{eq:action4}
\end{align}
Using the saddle point approximation we finally obtain the five coupled
equations of state 
\begin{eqnarray}
R & = & 2u'\mathbf{\Delta}_{0}^{2}+2u'\int_{p}{\rm tr}\left(\hat{{\cal \chi}}_{\mathbf{p}}\hat{\tau}_{0}\right)\label{eq:q0}\\
\mathbf{Q} & = & -2v\mathbf{\Delta}_{0}^{T}\hat{\boldsymbol{\tau}}\mathbf{\Delta}_{0}-2v\int_{p}{\rm tr}\left(\hat{{\cal \chi}}_{\mathbf{p}}\hat{\boldsymbol{\tau}}\right)\label{eq: q}\\
0 & = & 2\hat{{\cal \chi}}_{0}^{-1}\mathbf{\Delta}_{0}\;.\label{eq:Delta}
\end{eqnarray}
The saddle point value of the collective variable $\hat{Q}_{\alpha\beta}$
equals the desired order parameter $\left\langle \hat{q}_{\alpha\beta}\right\rangle $.
Vestigial nematicity is fluctuation induced. Without such superconducting
fluctuations, equation (\ref{eq: q}) does not allow for a finite
nematic order, $q_{0}\neq0$, above $T_{c}$.

\paragraph{Transition temperatures.}

The result of the numerical solution of the coupled set of equations
(\ref{eq:q0})-(\ref{eq:Delta}) are shown in Fig.\ref{fig:OP_fig}.
Here, we plot the order parameters as function of $\delta r=r_{0}-r_{0}^{*}\propto(T-T^{*}),$
where $T^{*}$ denotes the transition temperature without nematic
order present, i.e. $r_{0}^{*}=-2u'\int_{p}{\rm tr}\left(\hat{{\cal \chi}}_{p}\hat{\tau}_{0}\right)\Big|_{r=\mathbf{Q}=0}$.
We find that the nematic order parameter sets in above the superconducting
transition temperature $T_{c}$. The superconducting transition is
of second order and $T_{c}$ can also be obtained from the divergence
of the uniform pairing susceptibility $\chi_{{\rm pair}}^{-1}=(r^{2}-\mathbf{Q}^{2})/r\,$
with $r=r_{0}+R$, see Fig.\ref{fig:OP_fig}. Note that $\chi_{{\rm pair}}$
denotes the largest eigenvalue of the matrix $\hat{\chi}_{\mathbf{p}}$
for $\mathbf{p}=0$. The nematic transition at $T_{{\rm nem}}$ is
weakly first order. The origin of this behavior is the trigonal symmetry
which allows for a cubic invariant. Up to fourth order terms, the
action for the real order parameter $\boldsymbol{Q}$ reads:

\begin{align}
\mathcal{S}_{Q} & =\frac{r_{Q}}{2}\int_{x}(Q_{1}^{2}+Q_{2}^{2})+\frac{w}{3}\int_{x}Q_{1}(Q_{1}^{2}-3Q_{2}^{2})\nonumber \\
 & \quad+\frac{u_{Q}}{4}\int_{x}(Q_{1}^{2}+Q_{2}^{2})^{2}\;.\label{eq:nem_expansion-1}
\end{align}
This is the well-known Landau expansion of a three states Potts model
\cite{Straley1973}. Expanding the coupled set of equations (\ref{eq:q0})-(\ref{eq:Delta})
for small $Q_{i}$ yields exactly this term with $w\propto\bar{d}^{2}d'>0$.
\textcolor{black}{Thus, overall} the first order transition is expected
to be weak, \textcolor{black}{where it holds for the jump of the order
parameter} $\delta q_{0}\propto w$. Given the uncertainty in several
parameters we cannot reliably predict $T_{{\rm nem}}-T_{{\rm c}}$
in Kelvin. Our numerical analysis suggests however that it can be
up to $10\%$ of $T_{{\rm c}}$. In Fig.\ref{fig:OP_fig} we see also
the positive effect of nematic order on superconductivity. Fluctuations
without nematic order suppress the transition temperature to $T^{*}\ll T_{c}^{0}\approx\omega_{0}e^{-1/\lambda}$
as the coupling constant is reduced $\lambda^{-1}\rightarrow\lambda^{-1}+R/\rho_{F}$
. With nematic order this effect is significantly weakened as now
$\lambda^{-1}\rightarrow\lambda^{-1}+\left(R-\left|\mathbf{Q}\right|\right)/\rho_{F}$.
If we assume a more isotropic electronic structure we obtain instead
two joint first order transitions, a trend that also occurs in other
problems with vestigial precursor order \cite{Fernandes2012}. Photoemission
experiments \cite{Lahoud2012} for Cu$_{x}$Bi$_{2}$Se$_{3}$ strongly
support a very anisotropic Fermi surface, i.e. split transitions.

\paragraph{Degeneracy of superconducting and nematic ground states.}

\begin{figure}
\includegraphics[scale=0.4]{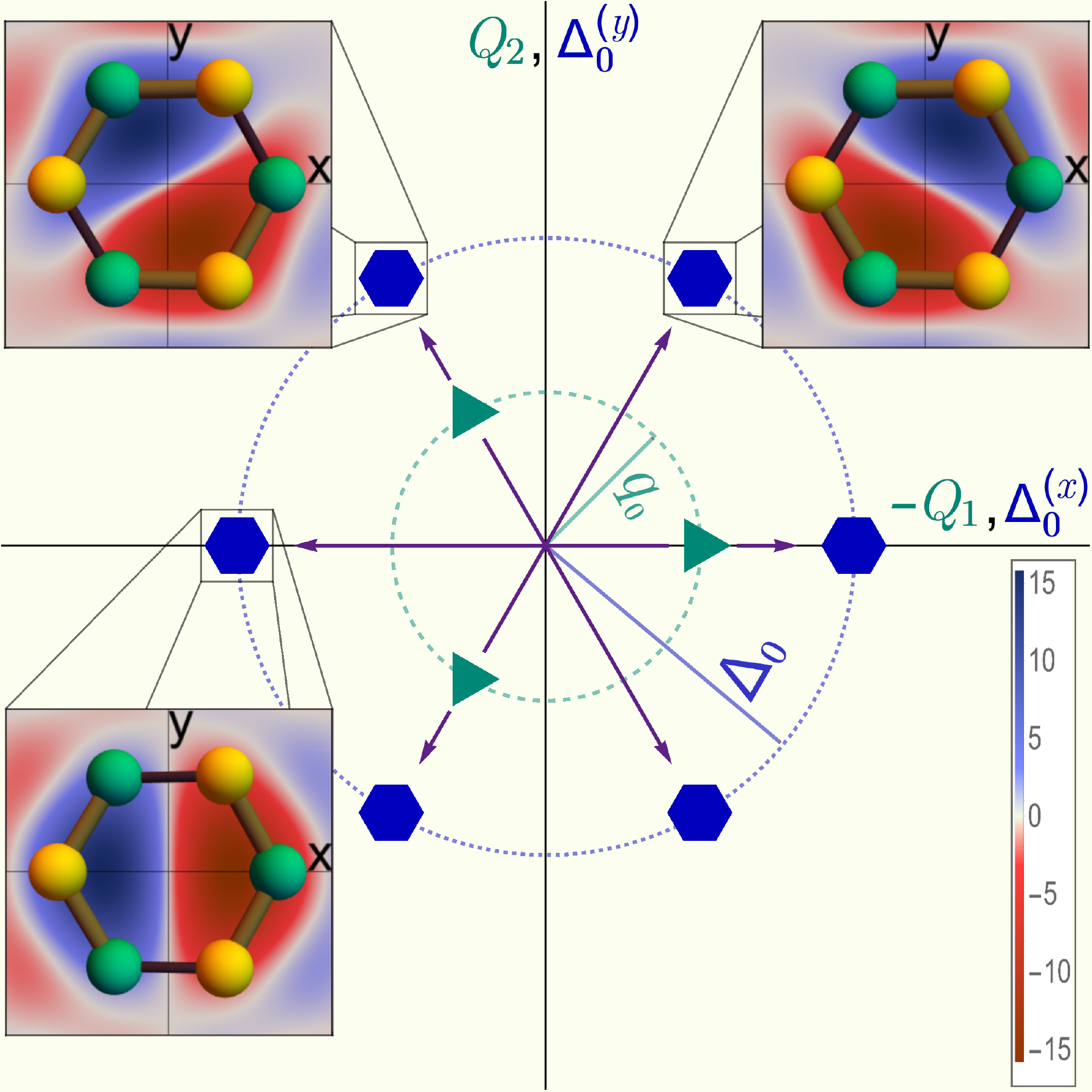}

\protect\protect\protect\caption{Illustration of the correspondence between the nematic and the superconducting
order parameters. The green triangles denote the threefold degenerate
nematic ground state (\ref{eq:nematic_solution}). Each nematic ground
state corresponds to two superconducting ground states (blue hexagons)
which differ by an overall phase of $\pi$ from one another, indicated
by the purple arrows. The three insets visualize the respective superconducting
ground state and the entailed threefold rotational symmetry breaking.
The $\mathbf{d}_{z}(x,y)$-component of the order parameter (see Eqn.(\ref{eq:d_vec}))
is plotted color coded on the ground with the hexagonal atomic structure
of the unit cell on top. The gap affects the electronic bonds differently,
leading to the aforementioned symmetry breaking.\label{fig:values_OPs}}
\end{figure}

Let us analyze the allowed orientation of the nematic director $\mathbf{n}$,
i.e. the allowed values for the angle $\theta$. As shown in Fig.\ref{fig:values_OPs},
the fluctuation induced term Eqn. (\ref{eq:nem_expansion-1}) picks
three distinct values of the angle $\theta=\left\{ 0,\frac{\pi}{3},\frac{2\pi}{3}\right\} $,
i.e. we have

\begin{align}
\mathbf{Q}= & q_{0}\,\left\{ \left(\begin{array}{c}
-1\\
0
\end{array}\right),\left(\begin{array}{c}
\frac{1}{2}\\
\frac{\sqrt{3}}{2}
\end{array}\right),\left(\begin{array}{c}
\frac{1}{2}\\
-\frac{\sqrt{3}}{2}
\end{array}\right)\right\} \;.\label{eq:nematic_solution}
\end{align}
The presence of a finite nematic order parameter predetermines which
of the degenerate superconducting ground states of Eqn. (\ref{eq:rot solution-1})
will be realized. At $T_{c}$ no additional rotational symmetry breaking
takes place and only the global $U\left(1\right)$ symmetry of the
superconductor and parity are broken. The one-to-one correspondence
between the superconducting and the nematic order parameters follows
from Eqn.(\ref{eq:Delta}), or equivalently, by determination of the
angle $\theta$ in (\ref{eq:nem_op}) and (\ref{eq:rot solution-1}).
We find that the first solution of (\ref{eq:nematic_solution}) leads
to the superconducting ground states with $\theta_{n}=\left\{ 0,\pi\right\} $,
while the second and third solutions lead to $\theta_{n}=\left\{ \frac{2\pi}{3},\frac{5\pi}{3}\right\} $
and $\theta_{n}=\left\{ \frac{\pi}{3},\frac{4\pi}{3}\right\} $, respectively
(see Fig.\ref{fig:values_OPs}). To visualize the origin of the in-plane
anisotropy in real space, the three insets of figure \ref{fig:values_OPs}
show the components of the triplet vector $\mathbf{d}_{z}(x,y)$(see
Eqn.(\ref{eq:d_vec})) of the respective superconducting ground states
in real space. We also show the Bi and Se atoms in the respective
layers of the crystalline unit cell to demonstrate how the bonds are
affected by the anisotropic superconducting gap. For the remainder
of the work, we choose, without restriction, the first of the three
degenerate nematic solutions, where $\theta_{n}=0$.

\paragraph{Experimental implications.}

In the following we study the experimental implication in the nematic
phase above $T_{c}$ and in the high temperature phase above $T_{{\rm nem}}.$
Above $T_{{\rm nem}}$ the onset of nematicity can be probed via renormalizations
of the elastic moduli of the system. The elastic energy relevant for
the transition is 
\begin{eqnarray}
\mathcal{S}_{{\rm el}} & = & \frac{1}{4}\int_{x}\, c_{A_{1g}}\,\left(\hat{\varepsilon}_{xx}+\hat{\varepsilon}_{yy}\right)^{2}\nonumber \\
 & + & \frac{1}{4}\int_{x}\, c_{E_{g}}\,\left[\left(\hat{\varepsilon}_{xx}-\hat{\varepsilon}_{yy}\right)^{2}+4\hat{\varepsilon}_{xy}\right],
\end{eqnarray}
where we focus on in-plane distortions. The symmetry allowed coupling
between the Cooper pair field and the elastic strain $\hat{\epsilon}_{\alpha\beta}$
is 
\begin{eqnarray}
\mathcal{S}_{{\rm nem-el}} & = & -\kappa\int_{x}{\rm tr}\left(\hat{\varepsilon}\hat{q}\right).
\end{eqnarray}
We can now add an external stress to the energy and determine the
renormalized elastic constants. Alternatively, we can add a conjugate
field to the nematic degrees of freedom and obtain the nematic susceptibility
\begin{equation}
\hat{\chi}_{{\rm nem},ij}\left(\mathbf{p}\right)=\left\langle Q_{i}\left(\mathbf{p}\right)Q_{j}\left(-\mathbf{p}\right)\right\rangle ,
\end{equation}
where the $Q_{i}$ are again the expansion parameters of the nematic
tensor in the Pauli basis $\hat{\boldsymbol{\tau}}=\left(\hat{\tau}^{z},\hat{\tau}^{x}\right)$
that we have been using. As long as the lattice is purely harmonic
we obtain the following relation between the renormalized elastic
modulus $c_{E_{g}}^{*}$ and its bare value $c_{E_{g}}$: 
\begin{equation}
(c_{E_{g}}^{*})^{-1}=(c_{E_{g}})^{-1}+\frac{\kappa^{2}}{2c_{E_{g}}}\,{\rm tr}\,\hat{\chi}_{{\rm nem}}\;,
\end{equation}
where $\hat{\chi}_{{\rm nem}}=\hat{\chi}_{{\rm nem}}\left(\mathbf{p}\rightarrow0\right)$.
A similar result for spin-induced nematicity was previously derived
in Ref.\cite{Fernandes2010}. As $T\rightarrow T_{{\rm nem}}$ from
above the nematic susceptibility rises, leading to a suppression of
elastic constants. Within the Gaussian fluctuation regime the nematic
susceptibility can be obtained explicitly and is given by: 
\begin{equation}
\hat{\chi}_{{\rm nem}}=2v\hat{\chi}_{{\rm nem}}^{\left(0\right)}\left(1-2v\hat{\chi}_{{\rm nem}}^{\left(0\right)}\right)^{-1},
\end{equation}
where $\hat{\chi}_{{\rm nem},ij}^{\left(0\right)}=\int_{p}{\rm tr}\left(\hat{\tau}_{i}\hat{\chi}_{\mathbf{p}}\hat{\tau}_{j}\hat{\chi}_{\mathbf{p}}\right)$.
In Fig.\ref{fig:OP_fig} we also show the temperature dependence of
$\hat{\chi}_{{\rm nem},ii}^{-1}$ which displays a Curie-Weiss dependence.
Since the transition is first order, $\hat{\chi}_{{\rm nem}}$ will
not diverge,\textcolor{black}{{} and for the offset holds} $\hat{\chi}_{{\rm nem}}^{-1}(T=T_{{\rm nem}})\propto w^{2}$.
However, as the first order transition is weak, the nematic susceptibility
is significantly enhanced. Note that $\hat{\chi}_{{\rm nem},ii}^{-1}\propto r_{Q}$
with $r_{Q}$ occuring in (\ref{eq:nem_expansion-1}) and $\hat{\chi}_{{\rm nem},ij}^{-1}=0$
for $i\neq j$ and $T>T_{{\rm nem}}$. To determine $\kappa$ and
the actual lattice softening one would need to know the change in
lattice parameters deep in the superconducting state. $\hat{\chi}_{{\rm nem}}$
is however directly observable via electronic Raman scattering \cite{Gallais2013,Kretzschmar2016}
in the $E_{g}$-channel.

\begin{figure}
\includegraphics[scale=0.4]{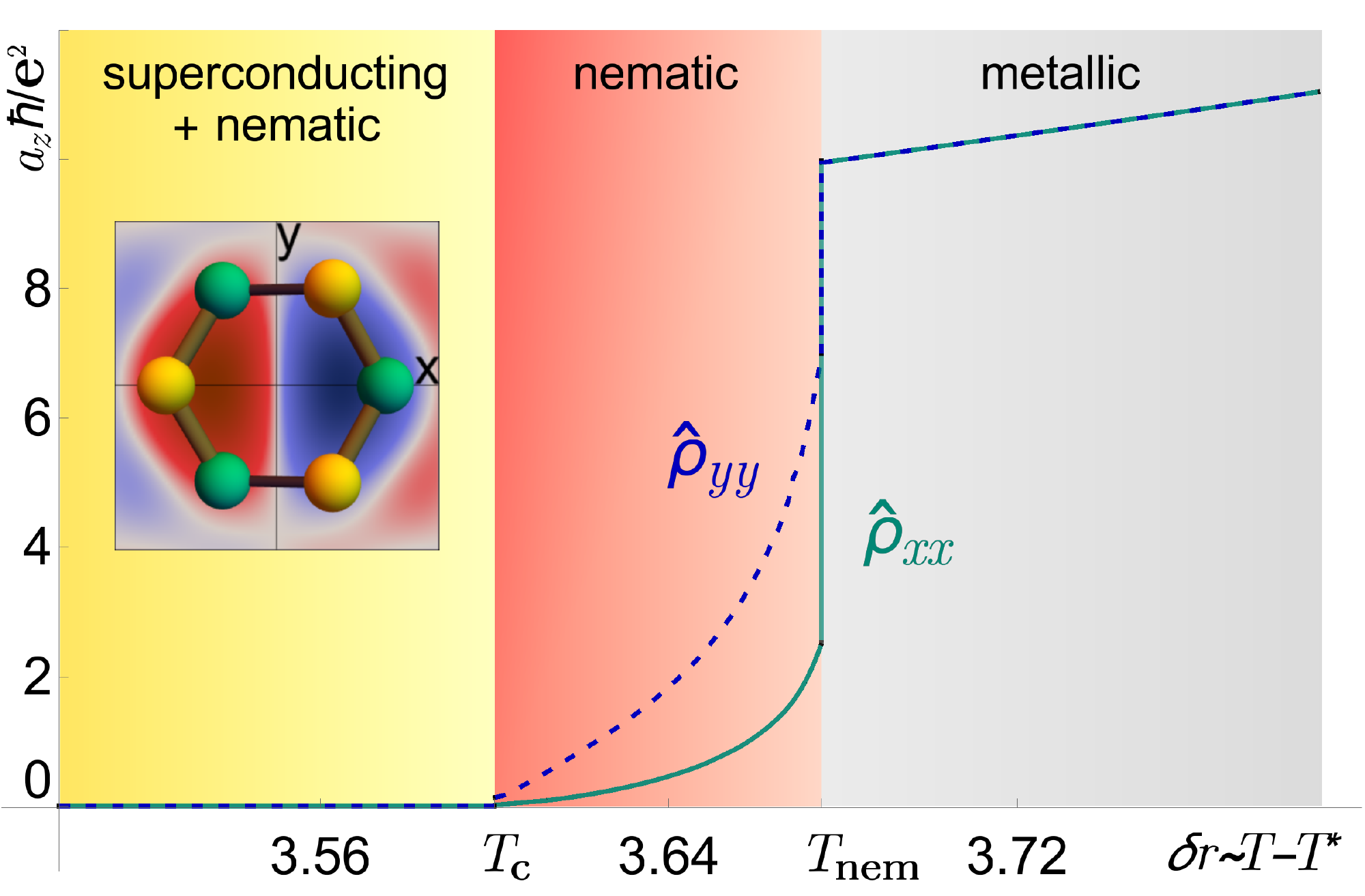}

\protect\protect\protect\caption{The calculated temperature behavior of the in-plane \textbf{dc-}resistivities,
$\hat{\rho}_{xx}$ (solid green) and $\hat{\rho}_{yy}$ (dashed blue).
The calculation only takes into account contributions from superconducting
order parameter fluctuations. As expected, we find an anisotropy in
the nematic phase where the sudden drop indicates that fluctuation
contributions become increasingly important in the nematic phase.
For the chosen ground state (depicted in the inset), the resistivity
in y-direction is larger, i.e. $\hat{\rho}_{yy}>\hat{\rho}_{xx}$.
\label{fig:resis}}
\end{figure}

Next, we study observables in the nematic phase, i.e. for $T_{c}<T<T_{{\rm nem}}$,
where the threefold symmetry is broken. As the nematic state is fluctuation
induced, the most natural quantity to reflect this anisotropy is the
paraconductivity of the system. Our calculation of the fluctuation
contribution to the resistivity is a natural generalization of the
classical works by Aslamasov and Larkin \cite{Azlamazov1968,Azlamazov1968-1}.
We obtain the conductivity:

\begin{align}
\hat{\sigma}_{\alpha\beta}\;\;\propto\;\;\frac{e^{2}a_{\alpha}a_{\beta}}{\hbar a_{x}a_{y}a_{z}}\;\times\;\;\vcenter{\hbox{\includegraphics[height=5em]{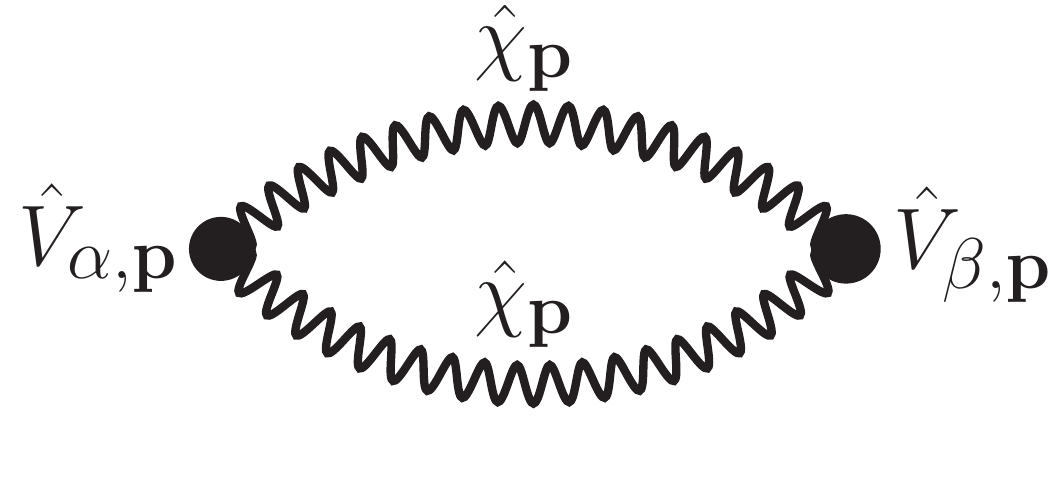}}} & \;,
\end{align}
with the lattice constants $a_{j}$, the velocity matrix $\hat{V}_{j,\mathbf{p}}=\partial\hat{\chi}_{\mathbf{p}}^{-1}/\partial p_{j}$
and the matrix of the pairing field $\hat{\chi}_{{\rm \mathbf{p}}}$.
The calculated temperature dependence of the resistivity $\hat{\rho}_{\alpha\alpha}=(\hat{\sigma}^{-1})_{\alpha\alpha}$
is plotted in figure \ref{fig:resis}. As expected, we find an anisotropy
between the two in-plane components. Moreover, the sudden drop at
$T_{{\rm nem}}$ once again indicates the first order nature of the
transition and evidences that fluctuation effects are more pronounced
inside the nematic phase. For the chosen ground state, i.e. $\theta_{n}=0$,
the resistivity in y-direction $\hat{\rho}_{yy}$ is larger than $\hat{\rho}_{xx}$,
since the fluctuating pairing amplitude along the $x$-direction is
much larger than in the orthogonal direction, see Fig.\ref{fig:values_OPs}.

\paragraph{Effect of disorder on the nematic phase.}

Apart from the usual pair-breaking effects, disorder has a profound
impact on states with vestigial order (see also \cite{Nie2014}).
A disorder configuration that locally changes a certain crystalline
orientation will naturally nucleate a specific value of the nematic
order parameter in its vicinity. Thus, ordinary potential scatters
act as random field disorder for the vestigial order parameter, which
is according to (\ref{eq:nem_expansion-1}) a three states Potts variable.
The random-field three-states Potts model was analyzed in \cite{Blankschtein1973,Eichhorn1996}.
Using the results of these papers the implication for our problem
is that disorder changes the first order transition to become second
order. Thus, the lattice softening should become more pronounced.
Most importantly, weak disorder is not expected to destroy the nematic
state.

\section*{Discussion}

We showed that superconductivity with odd-parity pairing in the doped
topological insulator Bi$_{2}$Se$_{3}$ is dramatically affected
by fluctuations. These fluctuations are important given the low carrier
concentration. As a result, the $U\left(1\right)$ gauge symmetry
and the rotational symmetry are separately broken at the temperatures
$T_{{\rm c}}$ and $T_{{\rm nem}}>T_{{\rm c}}$, respectively. The
intermediate nematic state is characterized by strong anisotropic
superconducting fluctuations that give rise to an anisotropic paraconductivity.
The symmetry breaking will certainly be inherited by the helical surface
states of Bi$_{2}$Se$_{3}$. The three states Potts universality
class of the vestigial order parameter implies that there should be
three distinct domains of vestigial order that can be aligned by applying
external stress in the $E_{g}$ symmetry, i.e. for finite $\hat{\epsilon}_{xx}-\hat{\epsilon}_{yy}$
or $\hat{\epsilon}_{xy}$. In addition, a lattice softening and enhanced
Raman response are expected above $T_{{\rm nem}}$. This mechanism
of composite order gives rise to an enhancement of the superconductivity
if compared to the usual fluctuation suppression of $T_{{\rm c}}$.
While the transition temperature should still be smaller than the
mean-field temperature, it offers an explanation for the comparatively
large transition temperature of Cu$_{x}$Bi$_{2}$Se$_{3}$, given
the low carrier concentration. This observation further suggests to
search for similar states of composite order in other low-carrier
superconductors with strong spin-orbit interaction.

\section*{Methods}

\paragraph{Derivation of Ginzburg-Landau expansion.}

We start from the established electronic structure of Bi$_{2}$Se$_{3}$
near the center of the Brillouin zone with the Hamiltonian \cite{Zhang09}
$H_{0}=\sum_{\mathbf{k}}c_{\mathbf{k}}^{\dagger}\left(h_{\mathbf{k}}-\mu_{0}\right)c_{\mathbf{k}}$,
where 
\begin{equation}
h_{\mathbf{k}}=v_{0}\hat{\tau}^{z}\left(k_{x}\hat{\sigma}^{y}-k_{y}\hat{\sigma}^{x}\right)+\left(v_{z}k_{z}\hat{\tau}^{y}+m\hat{\tau}^{x}\right)\hat{\sigma}^{0}+h_{\mathbf{k}}^{'}\,.\label{eq:h_k-1}
\end{equation}
Here, $c_{\mathbf{k}}=\left(c_{\mathbf{k},+\uparrow},c_{\mathbf{k},+\downarrow},c_{\mathbf{k},-\uparrow},c_{\mathbf{k},-\downarrow}\right)$
refers to the electron annihilation operators for momentum $\mathbf{k}$,
located in the two relevant $p_{z}$-orbitals in the unit cell ($\pm$),
and with spin ($\uparrow\downarrow$) \cite{Zhang09}. The Pauli matrices
$\hat{\tau}^{i}$ and $\hat{\sigma}^{j}$ act in orbital and spin
space, respectively. The last term $h_{\mathbf{k}}^{'}=-\lambda\left(k_{+}^{3}+k_{-}^{3}\right)\hat{\tau}^{z}\hat{\sigma}^{z}$
with $k_{\pm}=k_{x}\pm ik_{y}$ takes into account the point symmetry
of the hexagonal lattice. The origin of the Rashba-type spin-orbit
interaction in Eqn.(\ref{eq:h_k-1}) is caused by the lack of local
inversion symmetry of the Bi-Se layers. Globally, the system is inversion
symmetric, hence the coupling to $\hat{\tau}^{z}$, that is odd under
parity, in orbital space.

The superconducting pairing states of doped Bi$_{2}$Se$_{3}$ were
classified in Ref.\cite{Fu2010}. The state that is compatible with
a rotational symmetry breaking has odd-parity and gives rise to the
expectation value $\left\langle c_{\mathbf{k}}^{\dagger}\hat{\tau}^{y}\left(i\hat{\sigma}^{\mu}\hat{\sigma}^{y}\right)c_{-\mathbf{k}}^{\dagger}\right\rangle $,
where $\mu=\left\{ x,y\right\} $. Thus, we consider linear combinations
of equal spin pairing between distinct orbitals $\left\langle c_{\mathbf{k},+\uparrow}^{\dagger}c_{-\mathbf{k},-\uparrow}^{\dagger}\right\rangle $
and $\left\langle c_{\mathbf{k},+\downarrow}^{\dagger}c_{-\mathbf{k},-\downarrow}^{\dagger}\right\rangle $.
Such a state is generated by the Hamiltonian $H=H_{0}+H_{{\rm pair}}$
with pairing interaction 
\begin{equation}
H_{{\rm pair}}=-g\sum_{\mathbf{k,}\mathbf{k}',\mathbf{p},\mu=\left\{ x,y\right\} }b_{\mathbf{k},\mathbf{p}}^{\dagger\left(\mu\right)}b_{\mathbf{k'},\mathbf{p}}^{\left(\mu\right)}
\end{equation}
in the odd-pairity symmetry channel: $b_{\mathbf{k,}\mathbf{p}}^{\dagger\left(\mu\right)}=c_{\mathbf{k}}^{\dagger}\hat{\tau}^{y}\left(-i\hat{\sigma}^{\mu}\hat{\sigma}^{y}\right)c_{-\mathbf{k}+\mathbf{p}}^{\dagger}$.
The band structure of $H_{0}$ gives rise to four bands. In bulk,
two pairs of Kramers degenerate bands are separated by a gap. In electron
doped Bi$_{2}$Se$_{3}$, such as Cu$_{x}$Bi$_{2}$Se$_{3}$, the
Fermi energy is shifted to the upper two bands. Thus, we follow Ref.\cite{Venderbos2016}
and project into the conduction bands. Specifically, we use the manifestly
covariant Bloch basis $\psi_{\mathbf{k}s}$, see Ref.\cite{Fu15},
that respects the transformation behavior in coordinate and spin space.
The index $s=\left\{ 1,2\right\} $ refers to the pseudo-spin that
labels Kramers degeneracy. It follows $H_{0}\approx\sum_{\mathbf{k}s}\psi_{\mathbf{k},s}^{\dagger}\left(\varepsilon_{\mathbf{k}}-\mu_{0}\right)\psi_{\mathbf{k},s}$
with $\varepsilon_{\mathbf{k}}=\sqrt{m^{2}+v_{0}^{2}\left(k_{x}^{2}+k_{y}^{2}\right)+v_{z}^{2}k_{z}^{2}}$.
The pair creation operator in this basis is given by 
\begin{equation}
b_{\mathbf{k,}\mathbf{p}}^{\dagger\left(\mu\right)}=-\psi_{\mathbf{k}}^{\dagger}\left(\varphi_{\mathbf{k},\mathbf{p}}^{\left(\mu\right)}i\tilde{\sigma}^{y}\right)\psi_{-\mathbf{k}+\mathbf{p}}^{\dagger}\,,
\end{equation}
where $\psi_{\mathbf{k}}^{\dagger}=\left(\psi_{\mathbf{k},1}^{\dagger},\psi_{\mathbf{k},2}^{\dagger}\right)$
and the $\tilde{\sigma}{}^{l}$ are Pauli matrices in pseudo-spin
space. The full expressions for the form factors $\varphi_{\mathbf{k},\mathbf{p}}^{\left(\mu\right)}=d_{\mathbf{k},\mathbf{p}}^{s\left(\mu\right)}\tilde{\sigma}^{0}+\mathbf{d}_{\mathbf{k},\mathbf{p}}^{\left(\mu\right)}\cdot\boldsymbol{\tilde{\sigma}}$
are given by $d_{\mathbf{k},\mathbf{p}}^{s\left(x\right)}=-if_{\mathbf{k},\mathbf{p}}^{y,-}\,$,
$d_{\mathbf{k},\mathbf{p}}^{s\left(y\right)}=if_{\mathbf{k},\mathbf{p}}^{x,-}\,$,
$\mathbf{d}_{\mathbf{k},\mathbf{p}}^{\left(x\right)}=\left(f_{\mathbf{k},\mathbf{p}}^{z,+},0,-f_{\mathbf{k},\mathbf{p}}^{x,+}\right)$
and $\mathbf{d}_{\mathbf{k},\mathbf{p}}^{\left(y\right)}=\left(0,f_{\mathbf{k},\mathbf{p}}^{z,+},-f_{\mathbf{k},\mathbf{p}}^{y,+}\right)\;,$
where

\begin{align}
f_{\mathbf{k},\mathbf{p}}^{j,\pm} & =\frac{v_{j}}{2}\,\frac{(m+\epsilon_{\mathbf{k}})(k_{j}-p_{j})\pm(m+\epsilon_{\mathbf{k}-\mathbf{p}})k_{j}}{\sqrt{\epsilon_{\mathbf{k}}\epsilon_{\mathbf{k}-\mathbf{p}}(m+\epsilon_{\mathbf{k}})(m+\epsilon_{\mathbf{k}-\mathbf{p}})}}\;,\label{eq:d_vector}
\end{align}
$j=\left\{ x,y,z\right\} $, and $v_{x}=v_{y}=v_{0}$. For $\mathbf{p}\rightarrow\mathbf{0}$
holds that $d_{\mathbf{k},\mathbf{0}}^{s\left(\mu\right)}=0$ and
$\mathbf{d}_{\mathbf{k},0}^{\left(x\right)}=\frac{1}{\varepsilon_{\mathbf{k}}}\left(v_{z}k_{z},0,-v_{0}k_{x}\right)$
as well as $\mathbf{d}_{\mathbf{k},0}^{\left(y\right)}=\frac{1}{\varepsilon_{\mathbf{k}}}\left(0,v_{z}k_{z},-v_{0}k_{y}\right)$,
see also Ref.\cite{Venderbos2016}. This fully defines the low-energy
Hamiltonian of doped Bi$_{2}$Se$_{3}$. For the evaluation of the
Ginzburg-Landau parameters in Eqns. (\ref{eq:eq:action_delta}) and
(\ref{eq:gradient_action}) we use the parameters $v_{0}=3.3\,\text{eV}\,\text{\ensuremath{\mathring{A}}}$,
$m=0.28\,\text{eV}$ as in \cite{Zhang10}, $\mu=0.50\,\text{eV}$
as in \cite{Hashimoto2013}, $a\equiv a_{x}=a_{y}=4.1\,\text{\ensuremath{\mathring{A}}}$
as in \cite{isspLattConst} and $T_{c}=3.8\,\text{K}$. We chose $g$
to reproduce the experimental transition temperature. As there is
experimental evidence that the velocity $v_{z}$ depends on the amount
of Cu substitution, we kept $v_{z}$ as a tuning parameter. Depending
on the choice of $v_{z}$ we found either a joint first order transition,
or the scenario depicted e.g. in figure \ref{fig:OP_fig} where we
used $(v_{z}/a_{z})/(v_{0}/a)=1/20$. The plotted dimensionless quantities
from figure \ref{fig:OP_fig} read $\Delta_{0}\sqrt{d_{\parallel}V_{0}^{2}/(T_{c}V\tilde{v})}$,
$q_{0}/(2d_{\parallel}\tilde{v})$, $\chi_{{\rm pair}}^{-1}/(d_{\parallel}\tilde{v})$
and $\delta r/(d_{\parallel}\tilde{v})$, where we defined $\tilde{v}=(vT_{c})/(d_{\parallel}^{2}V_{0})$
and $V_{0}=a_{x}a_{y}a_{z}$. 

\begin{acknowledgments}
We are grateful to Yoichi Ando, Rafael M. Fernandes, Ian R. Fisher,
and Guo-qing Zheng for helpful discussions. The work of J.S. was performed
in part at the Aspen Center for Physics, which is supported by NSF
grant PHY-1607611.
\end{acknowledgments}

\section*{Competing interests}

The authors declare that they have no competing interests.

\section*{Author Contributions}

M.H. and J.S. performed the research and wrote the paper.

\section*{Data Availability}

The data that support the findings of this study are available from
the authors on request.
\end{document}